\documentclass[aps,prx,superscriptaddress,twocolumn,amssymb,amsmath,longbibliography,floatfix]{revtex4-2}

\usepackage{graphicx}
\usepackage{amsmath,amssymb,bm}
\usepackage{xcolor}
\usepackage[colorlinks,urlcolor=blue,citecolor=blue,linkcolor=blue,filecolor=blue,unicode=true]{hyperref}
\usepackage{multibib}
\newcommand{\micron}{\ensuremath{\mu\mathrm{m}}}

\newcommand{\fivevec}[5]{\left(\begin{array}{c}#1\\#2\\#3\\#4\\#5\end{array}\right)}

\newcommand{\ket}[1]{\ensuremath{\left|#1\right>}}

\newcommand{\SO}{\ensuremath{\mathrm{SO}}}
\newcommand{\SU}{\ensuremath{\mathrm{SU}}}
\newcommand{\D}{\ensuremath{\mathrm{D}}}

\newcommand{\Dt}{\ensuremath{\tilde{\D}}}
\newcommand{\sx}{\ensuremath{\sigma_x}}
\newcommand{\sy}{\ensuremath{\sigma_y}}
\newcommand{\sz}{\ensuremath{\sigma_z}}

\newcommand{\Fhat}{\ensuremath{\mathbf{\hat{F}}}}
\newcommand{\rr}{\ensuremath{\mathbf{r}}}
\newcommand{\nuke}[1]{}

\newcommand{\beq}{\begin{equation}}
\newcommand{\eeq}{\end{equation}}

\begin{document}

\title{Topological Superfluid Defects with Discrete Point Group Symmetries}

\author{Y. Xiao}
\altaffiliation{Current address: Department of Electrical Engineering and Computer Science, University of Michigan, Ann Arbor, Michigan 48109, USA}
\affiliation{Department of Physics and Astronomy, Amherst College, Amherst, Massachusetts 01002--5000, USA}

\author{M.~O. Borgh}
\affiliation{Physics, Faculty of Science, University of East Anglia, Norwich, NR4 7TJ, United Kingdom}

\author{A.~A. Blinova}
\affiliation{Department of Physics, University of Massachusetts Amherst, Amherst, Massachusetts 01003, USA}
\affiliation{Department of Physics and Astronomy, Amherst College, Amherst, Massachusetts 01002--5000, USA}

\author{T.~Ollikainen}
\altaffiliation{Current address: Institut f\"ur Quantenoptik und Quanteninformation, \"Osterreichische Akademie der Wissenschaften, Technikerstra\ss e 21a, 6020 Innsbruck, Austria.}
\affiliation{QCD Labs, QTF Centre of Excellence, Department of Applied Physics, Aalto University, P.O. Box 13500, FI--00076 Aalto, Finland}
\affiliation{Department of Physics and Astronomy, Amherst College, Amherst, Massachusetts 01002--5000, USA}

\author{J. Ruostekoski}
\affiliation{Department of Physics, Lancaster University, Lancaster, LA1 4YB, United Kingdom}

\author{D.~S. Hall}
\affiliation{Department of Physics and Astronomy, Amherst College, Amherst, Massachusetts 01002--5000, USA}

\date{\today}

\begin{abstract}
Discrete symmetries are spatially ubiquitous but are often hidden in internal states of systems where they can have especially profound consequences. In this work we create and verify exotic magnetic phases of atomic spinor Bose-Einstein condensates that, despite their continuous character and intrinsic spatial isotropy, exhibit complex discrete polytope symmetries in their topological defects. Using carefully tailored spinor rotations and microwave transitions, we engineer singular line defects whose quantization conditions, exchange statistics, and dynamics are fundamentally determined by these underlying symmetries. We show how filling the vortex line singularities with atoms in a variety of different phases leads to core structures that possess magnetic interfaces with rich combinations of discrete and continuous symmetries. Such defects, with their non-commutative properties, could provide unconventional realizations of quantum information and interferometry.
\end{abstract}

\maketitle

\pagebreak

\section{INTRODUCTION}

Symmetry plays a critical role in the scientific and mathematical descriptions of the universe. Symmetries can be continuous, as in rotations of a circular cylinder about its axis; or discrete, as in end-for-end exchanges of the cylinder about its midpoint. Discrete polytope symmetries appear in diverse and widespread systems, including crystals, molecular bonds, and the familiar morphologies of honeycombs, snowflakes, and flower petals. They can also be hidden in the internal states of otherwise continuous and isotropic systems, where they can have profound and unusual consequences;
for example, the discrete symmetries of charge conjugation, parity, and time-reversal play important roles in particle and condensed matter physics and serve as a touchstone for grand unified theories. Complex discrete symmetries also appear in spatially uniform condensed matter systems and, intriguingly, in spinor superfluids.

\begin{figure*}[bht!]
\centering
\includegraphics[width=\linewidth]{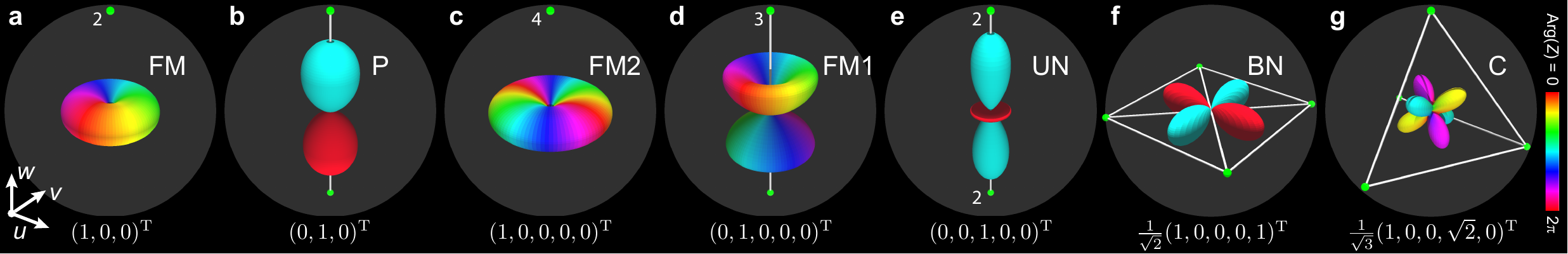}
\caption{\textbf{Majorana and spherical-harmonics representations of the prototype spinors for spin-1 and spin-2 magnetic phases.}
\textbf{a,b,} The spin-1 ferromagnetic (FM) and polar (P) magnetic phases with two Majorana points (green dots, with adjacent number indicating multiplicity $>1$). \textbf{c--g,} The spin-2 ferromagnetic-2 (FM2) and -1 (FM1),  uniaxial nematic (UN), biaxial nematic (BN), and cyclic (C) magnetic phases, with four Majorana points. The discrete polytope Majorana symmetries of a square and tetrahedron are easily recognized for BN and C. The full behavior of the order-parameter symmetries is visualized in the spherical harmonics representation, where $Z(\theta,\phi)$, for spherical coordinates $(\theta,\phi)$, expands each spinor in terms of spherical harmonics. The shape $|Z(\theta,\phi)|^2$ and $\mathrm{Arg}(Z)$ (color map) together reveal the symmetry.
 The FM, FM1 and FM2 order parameters correspond to spatial rotations in three dimensions.
 The order parameter symmetries of the remaining magnetic phases are obtained by appropriately combining the global condensate phase with an unoriented axis (P and UN), square (BN), and tetrahedron (C).
}
\label{fig:phases}
\end{figure*}

In quantum mechanics, the internal symmetries of a spin-$F$ system can be conveniently described in a geometrical representation due to Majorana, wherein a state corresponds to a constellation of $2F$ points on the unit sphere~\cite{majorana_nuovocimento_1932,bloch_rmp_1945}. Each point is related to the state of an independent spin-1/2 system~\cite{schwinger_book_1965}, and the polytopes with vertices established by the representative points display the discrete symmetries of the order parameter that describes the system~\cite{bacry_jmathphys_1974,barnett_prl_2006,makela_prl_2007}. The stationary, dynamically stable states that share a given constellation are known as magnetic phases, whose richness of internal symmetries is illustrated by the examples in  Fig.~\ref{fig:phases}.

In contrast to the ubiquity of their crystalline counterparts, condensed matter systems exhibiting discrete polytope symmetries within their internal degrees of freedom are relatively unusual, with examples appearing in exotic contexts such as $d$-wave superconductors~\cite{mermin_pra_1974}, $^3\mathrm{P}_2$ neutron star superfluids~\cite{hoffberg_prl_1970}, and biaxial nematic liquid crystals~\cite{poenaru_jphys_1977}. Spinor Bose--Einstein condensates (BECs)~\cite{kawaguchi_physrep_2012} with spin $F\geq 2$ provide an exciting pristine system with unprecedented experimental control. They are described by an order parameter whose complex broken internal symmetries lead to magnetic phases with polygonal (e.g., Fig.~\ref{fig:phases}f), tetrahedral (e.g., Fig.~\ref{fig:phases}g), octahedral  ($F\geq 3$), and icosahedral ($F\geq 6$) Majorana symmetries. Previous experimental studies have examined phenomena associated with spinor BECs in simple magnetic phases that lack such symmetries, including topological defects~\cite{seo_prl_2015,ray_science_2015,weiss_ncomm_2019,xiao_commphys_2021} and textures~\cite{leanhardt_prl_2003,leslie_prl_2009,choi_prl_2012,lee_sciadv_2018}, spontaneous pattern formation~\cite{sadler_nature_2006,scherer_prl_2010}, dipolar interactions~\cite{lepoutre_prl_2018}, magnons~\cite{marti_prl_2014}, and condensate fragmentation~\cite{evrard_science_2021}.
However, apart from early experiments on spin population dynamics~\cite{schmaljohann_prl_2004}, magnetic phases with discrete polytope symmetries remain unexplored.

Here, we create and verify the discrete polytope symmetries of biaxial nematic (BN) and cyclic (C) phases of spin-2 BECs in the rotational properties of their continuous wave functions and in the singular line defects. The BN and C magnetic phases exhibit discrete internal Majorana symmetries of a square (Fig.~\ref{fig:phases}f) and a tetrahedron (Fig.~\ref{fig:phases}g), respectively, which are revealed experimentally. Such phases support exotic singular vortices that we prepare using combinations of magnetic field rotations and carefully engineered microwave transitions. The singular defect cores 
are filled by a variety of different superfluid phases, resulting in rich combinations of core structures that form interfaces between different discrete and continuous symmetries. Our experimental procedure thus establishes a fascinating setting for the exploration and manipulation of the unusual properties of the line defects~\cite{semenoff_prl_2007,barnett_pra_2007,borgh_prl_2016} that directly emerge from the discrete polytope symmetries. For instance, the line defects may not commute when the vortex positions are interchanged, leaving rung vortices behind in collisions~\cite{kobayashi_prl_2009,borgh_prl_2016} with possible ramifications for interferometry and quantum information~\cite{mawson_prl_2019}.

\section{LINE DEFECTS WITH POLYTOPE SYMMETRIES}

We create line defects by applying a carefully tailored time- and spatially varying magnetic field to a spin-1 $^{87}$Rb superfluid in its ferromagnetic (FM) phase, which maximizes the spin magnitude, with the order parameter defined by spatial rotations (Fig~\ref{fig:phases}a).
The changing magnetic field imprints a nonsingular vortex texture with a tight bending of the magnetization in the vicinity of its core, triggering an instability that induces the system to decay into a pair of singly-quantized, singular vortices in the FM phase~\cite{lovegrove_pra_2016,weiss_ncomm_2019}. During the decay process the vortex cores fill with superfluid in the polar (P) magnetic phase, with nematic symmetry (Fig.~\ref{fig:phases}b) and vanishing condensate spin magnitude. The magnetic spin-2 superfluid phases are subsequently introduced with a sequence of microwave pulses that promote the spin-1 atoms into the spin-2 hyperfine manifold. The result is a pair of line defects in any of the prototypical spin-2 magnetic phases shown in Fig.~\ref{fig:phases}c--g, but we focus here on converting the FM phase to either the BN or C phase that display discrete polytope symmetries. In the same pulse sequence, the filled core in the P phase is similarly converted to a distinct $F=2$ magnetic phase of our choice, such as ferromagnetic-2 (FM2), with maximal spin magnitude and symmetry related to the spatial rotations, or uniaxial nematic (UN), with vanishing condensate spin and the symmetry defined by an unoriented axis and a $2\pi$ change of the global BEC phase. (Further details of the magnetic phases are provided in the Appendix.)

\begin{figure*}
\centering
\includegraphics[width=\linewidth]{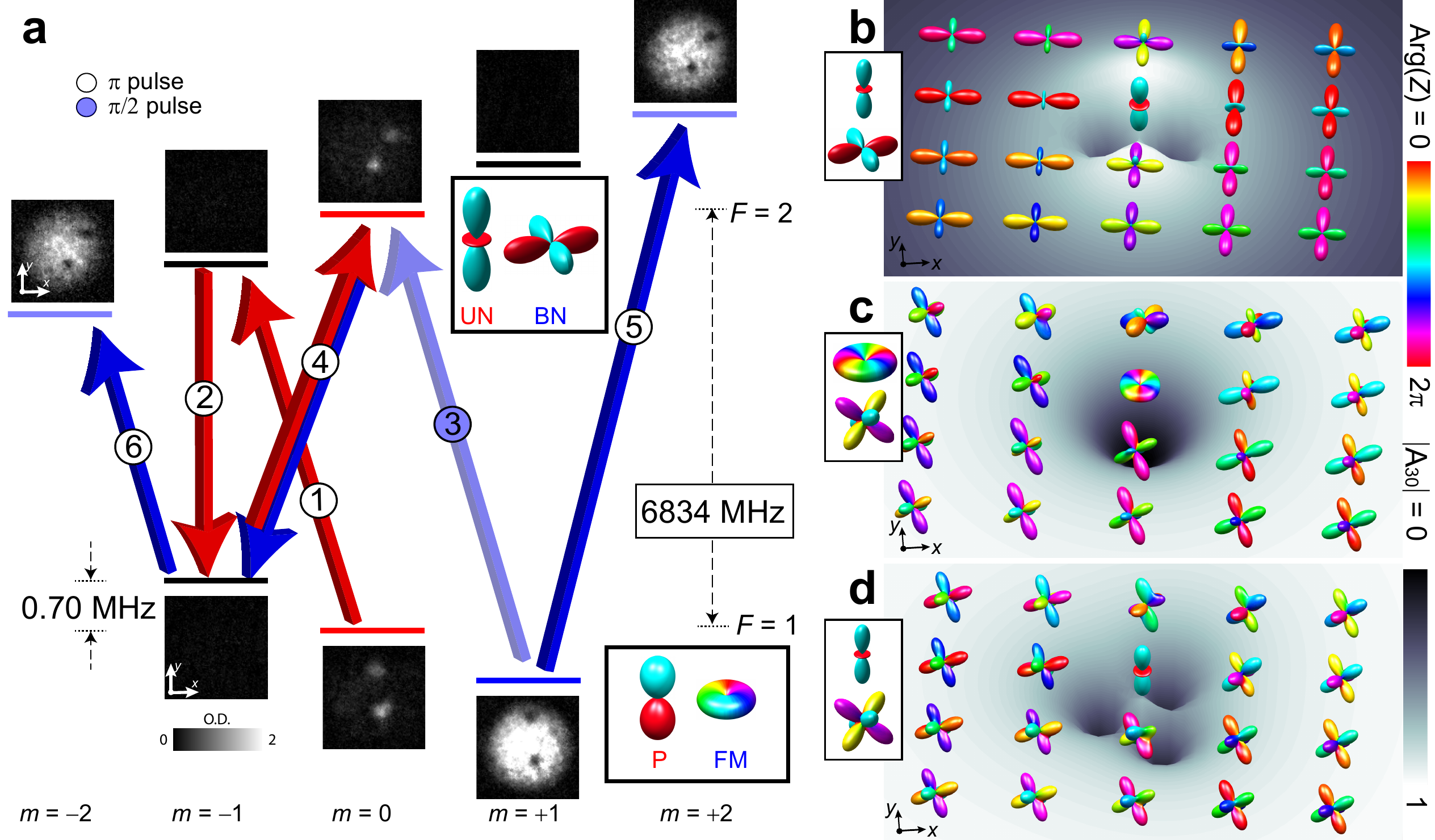}
\caption{\textbf{Creation of singular line defects in magnetic phases with discrete polytope symmetries.} \textbf{a,} Engineering a singular vortex in the BN phase with a UN vortex core. The thick lines schematically show the hyperfine levels ($F=1$ and $F=2$, with Zeeman levels $m$ reading left to right) in a magnetic field of 1~G, accompanied by experimental images of the condensate viewed along the $z$-axis after Stern--Gerlach separation. The connecting arrows illustrate the pulse sequence, with order given by the circled number, colored blue for transitions involving the components with phase singularities and red involving the superfluid components filling those singularities. Deep colors indicate $\pi$-pulses, whereas pale blue indicates a $\pi/2$-pulse.
The process begins with a vortex in the spin-1 FM phase with P core.
\textbf{b--d,} Spherical harmonics representation of the core structure of imprinted singly quantized vortices: \textbf{b,} BN vortex with UN core, corresponding to a single vortex as created in \textbf{a}; \textbf{c,} C vortex with ferromagnetic-2 core; \textbf{d,} C vortex with UN core. The symmetry of the order parameter is represented by the spherical harmonics expansion given in Fig.~\protect\ref{fig:phases}.
The background surface shows the singlet-trio amplitude $| A_{30}|$ (Appendix) that reveals a non-axisymmetrically connecting discrete magnetic phase at the topological interface between the vortex core and bulk superfluid. The column densities of the experimental images are expressed in grayscale in terms of dimensionless optical depth (O.D.), and the field of view of each image is $219~\micron \times 219~\micron$.
}
\label{fig:leveldiagram}
\end{figure*}

Our experimental creation technique accesses an entire family of line defects in which the polytope symmetries affect the properties of the vortex cores.
An example of a microwave pulse sequence that yields a pair of BN vortices with filled UN cores is shown in Fig.~\ref{fig:leveldiagram}a. In the chosen basis the BN condensate is represented by
a square in the $uv$- (horizontal) plane of the Majorana sphere (Fig~\ref{fig:phases}f), whereas the UN core is similarly represented by 
a rod oriented along the $w$- (vertical) axis (Fig~\ref{fig:phases}E).  More generally, changing the relative phase between the nonzero spinor components of either the BN or C phases (Fig~\ref{fig:phases}f,g) amounts to rotating the corresponding polytope about the $w$-axis, and is achieved experimentally by selecting the appropriate phases of the microwave pulses.

The formation of the vortex core can be understood in this example of
the singly-quantized BN line defect as the continuous transformation of the BN bulk phase towards the UN phase inside the vortex core (Fig.~\ref{fig:leveldiagram}b). We can parameterize the stationary spinor solution by
\begin{equation}
\label{eq:core}
\zeta = \fivevec{e^{i \phi}  f(\rho)/\sqrt{2}}{0}{\sqrt{1-[f(\rho)]^2}}{0}{e^{i \phi} f(\rho)/\sqrt{2}}\,,
\end{equation}
where $f(\rho)$ parameterizes the radial vortex-core profile in terms of the radial coordinate $\rho=(x^2+y^2)^{1/2}$, such that $f(\rho)\rightarrow 1$ outside the core, and $f(\rho)=0$ on the line singularity itself $(\rho= 0)$. This solution smoothly interpolates between the BN and UN phases. Hence, the total superfluid density remains non-zero as the vortex line is filled with the UN phase, even though the BN phase becomes singular and must vanish. This is unlike a defect singularity in a scalar superfluid that has a vanishing superfluid density. The solution given by Eq.~\ref{eq:core} thus constitutes a stable, coherent topological interface  at the vortex core~\cite{borgh_prl_2012,lovegrove_pra_2016}. More generally, such interfaces can support non-trivial topological objects whose classification changes as they penetrate the interface. These have been extensively studied, e.g., in high-energy physics and cosmology~\cite{kibble_jpa_1976,sarangi_plb_2002}, in superfluid liquid $^3$He~\cite{bradley_nphys_2008}, and
in BECs~\cite{borgh_prl_2012}.

Further instructive examples are illustrated in Fig.~\ref{fig:leveldiagram}c,d, in which the core of a C line defect is filled with either the FM2 or the UN phase (see also the Supplemental Material~\cite{supplementalmaterial}) for examples of these experimental realizations).
The Majorana representation of the C phase in these cases yields a vertex-up tetrahedron with a face parallel to the $uv$-plane. While the FM2 core is rotationally symmetric, the core of the resulting singly-quantized C vortex is remarkably anisotropic when the UN phase occupies the line singularity.
Numerical simulations show that the vortices in both cases are unstable against splitting into a pair of vortices carrying fractional $\tfrac{1}{3}$ and $\tfrac{2}{3}$ circulation quanta, respectively. In the process, the $\tfrac{2}{3}$ vortex develops a FM2 core, while the core of the $\tfrac{1}{3}$ vortex exhibits the ferromagnetic-1 (FM1) phase (cf., Fig.~\ref{fig:phases}c,d). However, the decay is sensitive to the orientation of the order parameter of the imprinted vortex. In simulations, we find that a $\pi/4$ spin rotation about the $v$ axis instead causes a decay of the condensate into the BN phase, forming vortices with FM2 cores.

Images of the condensate spinor components reveal the rich structure of line defects in the magnetic phases with polytope symmetries but do not directly show the discrete internal symmetries of the phases in real space. To verify these symmetries we apply radio-frequency (rf) spin-tip pulses that rotate the entire Majorana constellation about an axis in the $uv$-plane~\cite{majorana_nuovocimento_1932}, thereby preserving the magnetic phase but changing its spinor representation. The rotation axis and angle of rotation are established by the phase and area of the rf pulse, respectively, in the same fashion as they act on states on the Bloch sphere~\cite{bloch_rmp_1945}.
Critically, the spinor component densities remain invariant under any $2\pi/j$ adjustments of the rf phase along a $j$-fold symmetry axis of the polytope.

\begin{figure*}[ht!]
\centering
\includegraphics[width=0.85\linewidth]{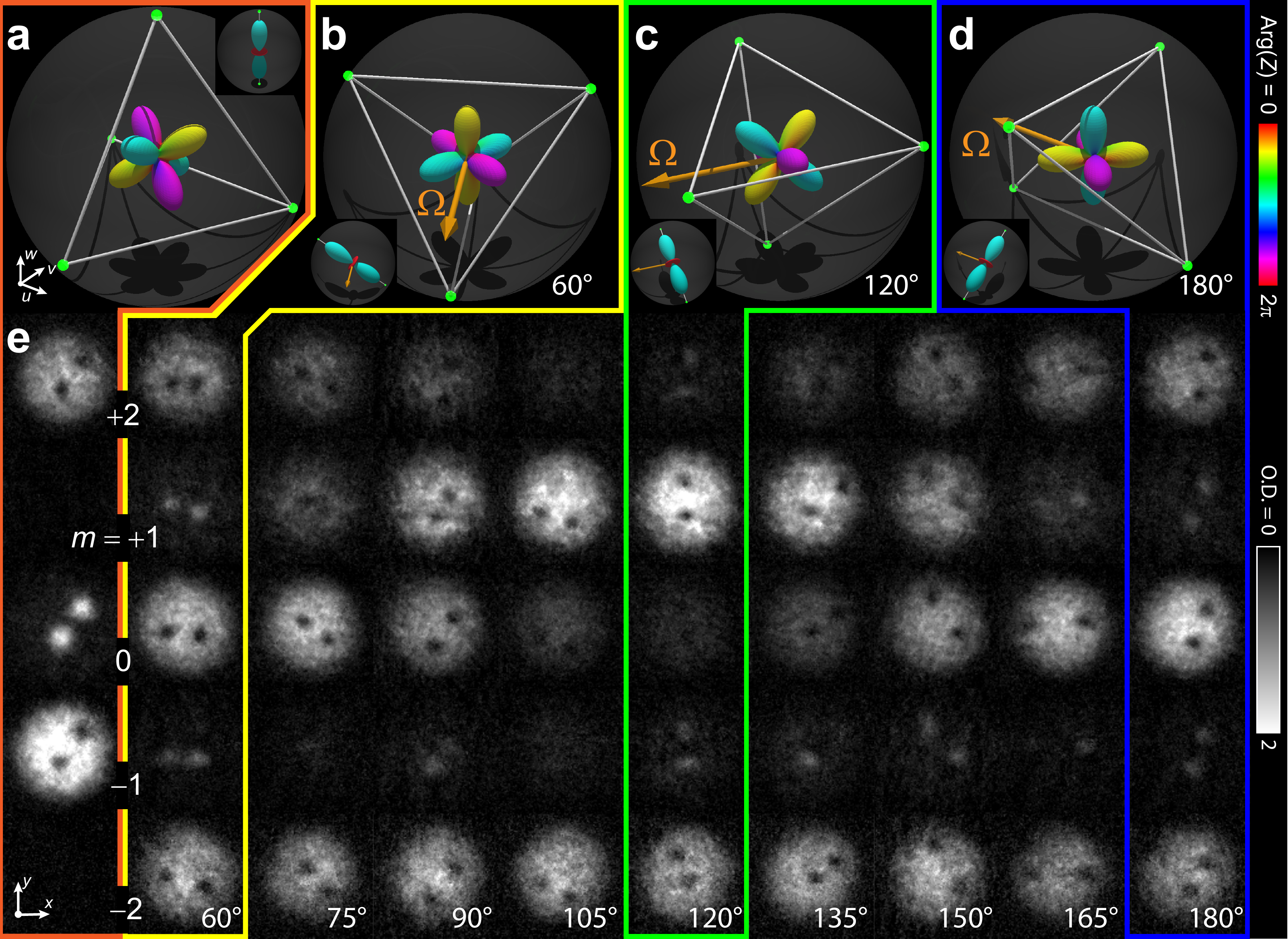}
\caption{\textbf{
Demonstration of tetrahedral discrete symmetry in the continuous wave function of the spin-2 superfluid in the C phase with a singular line defect.}  The orientation of the order parameter is controlled using rf pulses of varying phase. \textbf{a,} The initial C vortex state in the Majorana representation. The symmetry of the order parameter is represented by the spherical harmonics expansion given in Fig.~\protect\ref{fig:phases}.
The vortex core is in the UN phase whose initial state is shown in the inset. \textbf{b--d,} The C vortex states after application of a spin-tip pulse with relative phase as indicated and with torque vector $\bm{\Omega}$, represented in the rotating frame (with respect to axes $u$, $v$, and~$w$). The insets show the corresponding orientations of the uniaxial nematic core. \textbf{e,} The spinor component densities associated with the spin-tip pulses with the relative phases as shown. Each experimental spinor component image is $219~\micron \times 219~\micron$ and shows column density expressed in grayscale in terms of optical depth (OD).}
\label{fig:un_phases}
\end{figure*}

Extending our previous example, we demonstrate the three-fold discrete symmetry of the cyclic magnetic phase by rotating the vertex-up tetrahedral state of the C phase (Fig.~\ref{fig:phases}g) through the tetrahedral angle $\arccos(1/\sqrt{3})$ about the axis in the $uv$-plane established by the phase of the spin-tip pulse.
As the rf phase is varied through $\tfrac{2\pi}{3}$ radians (Fig.~\ref{fig:un_phases}b--e), the final vortex state undergoes one corresponding revolution that returns to the edge-up state $\tfrac{1}{2}(1,0,i\sqrt{2},0,1)^T$, as shown in Fig.~\ref{fig:un_phases}b,d. The imaginary amplitude of the $m=0$ component, inferred from the images after the rotation, is critical to distinguishing the cyclic magnetic phase from a mixed UN-BN phase that has the same spinor component densities but does not share the tetrahedral Majorana symmetry. Similar measurements reveal the four-fold Majorana symmetry of the BN phase under $\pi/2$ rotations (Fig.~\ref{fig:bnphase}) and the full three-fold symmetry of the C phase under $2\pi/3$ rotations (Fig.~\ref{fig:cphase}).

\section{CONCLUDING REMARKS}

\begin{figure}[ht!]
\centering
\includegraphics[width=\linewidth]{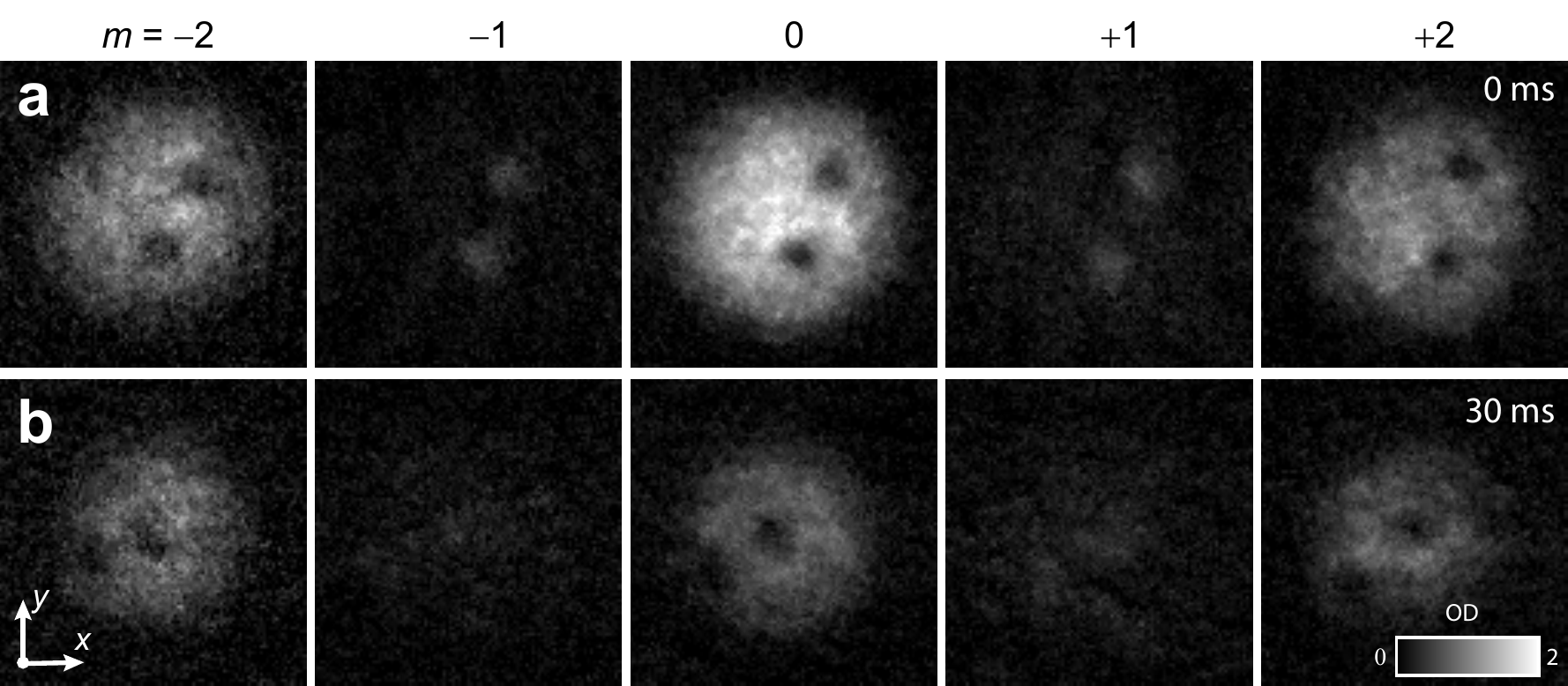}
\caption{\textbf{Initial time evolution of the singular line defects in the C phase, with the defect core in the UN phase.}  Each row shows the spinor components at an evolution time of \textbf{a,} 0~ms, and \textbf{b,} 30~ms after an applied tetrahedral angle rf pulse. Each experimental spinor component image is $219~\micron \times 219~\micron$, with column densities expressed in grayscale in terms of optical depth (O.D.).}
\label{fig:un-c_time}
\end{figure}

Our creation of line defects in magnetic phases with demonstrated polytope internal symmetries suggests a number of future experiments. As a proof of principle, we experimentally explored the time evolution of a UN core, C vortex condensate for the first 30~ms after a tetrahedral rf rotation into the edge-up tetrahedral state (Figs.~\ref{fig:un_phases}b,d and~\ref{fig:un-c_time}). Over this interval, the total number of atoms decreases substantially, whereas the fraction of atoms in each spinor component changes only slightly: the $m=\pm 1$ spinor components become more diffuse in the region of the vortex core, and the fraction of atoms in the $m=0$ component diminishes slightly with respect to the $m=\pm 2$ spinor components. These results suggest that the C superfluid trends slowly towards the BN phase, consistent with previous experimental studies~\cite{schmaljohann_prl_2004}. The relatively short superfluid lifetime is a serious constraint that can be addressed with the development of optical trapping geometries that reduce the atomic density. Overcoming this technical hurdle will establish a path to exploring the vortex dynamics more fully, including the expected evolution towards fractionally-quantized vortices we highlighted above.

Our engineering of magnetic phases with discrete polytope symmetries and the associated line defects also opens new avenues for the further exploration of exciting phenomena that have previously only been associated with exotic models in field theories of high-energy physics and cosmology. For instance, defects across the interfaces of polytope symmetric magnetic phases can combine non-trivially, forming singularities that penetrate continuous, coherent topological interfaces and non-Abelian line defects appear as cosmic strings in theories of the early Universe~\cite{vilenkin-shellard}. Non-Abelian defects in the magnetic phases of discrete polytope symmetries may furthermore form a non-commutative algebra~\cite{semenoff_prl_2007,borgh_prl_2016} that could potentially be harnessed for quantum information processing~\cite{mawson_prl_2019}. Experiments for manipulating such defects are challenging but may be assisted by working in, e.g., strongly oblate trapping geometries where the defects become more easily controllable point vortices.

\begin{acknowledgments}
The authors are grateful for experimental assistance from L.~Reuter in obtaining the data presented in Fig.~\ref{fig:bnphase}. D.S.H.\ and T.O.\ thank M.\ M\"ott\"onen for helpful discussions. D.S.H.\ acknowledges financial support from  the National Science Foundation (Grant No.\ PHY--1806318), J.R.\ from the UK EPSRC (Grant Nos.\ EP/P026133/1, EP/S002952/1), and T.O.\ from the Emil Aaltonen Foundation and the Kaupallisten ja teknillisten tieteiden tukis\"a\"ati\"o (KAUTE) foundation through its Researchers Abroad program.
\end{acknowledgments}

\appendix

\section{Spin-2 mean-field theory}

\subsection{Spin-2 Hamiltonian}
We represent the macroscopic spin-2 BEC wave function as $\Psi(\mathbf{r},t)=\sqrt{n(\mathbf{r},t)} e^{i\tau}\zeta$, such that the atom density $n=|\Psi|^2$, the global phase $\tau(\mathbf{r},t)$, and the spinor $\zeta(\mathbf{r},t)$ are governed by a mean-field Hamiltonian density~\cite{kawaguchi_physrep_2012}
\begin{align}
\begin{split}
\mathcal{H}= &\frac{\hbar^2}{2M_\text{a}}|\nabla\Psi|^2+U_\mathrm{trap}n\\ &+\frac{c_0}{2}n^2 + \frac{c_1}{2}n^2|\langle\Fhat\rangle|^2 + \frac{c_2}{2}n^2|A_{20}|^2\\ &+pn\langle\hat{F}_z\rangle+qn\langle\hat{F}_z^2\rangle
\end{split}
\label{ham}
\end{align}
in the harmonic trapping potential $U_\mathrm{trap}=(M_\text{a}\omega_r^2/2)(x^2+y^2+2z^2)$, with radial trap frequency $\omega_r$. Here $M_\text{a}$ denotes the atomic mass, $A_{20}=\frac{1}{\sqrt{5}}(2\zeta_{+2}\zeta_{-2}-2\zeta_{+1}\zeta_{-1}+\zeta_0^2)$ is the amplitude of spin-singlet pair formation, and $p$ and $q$ determine, respectively, the linear and quadratic Zeeman shifts. The condensate-spin expectation value $\langle\Fhat\rangle=\sum_{\alpha\beta} \zeta_\alpha^\dagger\Fhat_{\alpha\beta} \zeta_\beta$ is obtained from
 the vector $\Fhat$ of spin-2 matrices. The nonlinearities  $c_0=4\pi\hbar^2(3a_4+4a_2)/7M_\text{a}$, $c_1=4\pi\hbar^2(a_4-a_2)/7M_\text{a}$, and $c_2=4\pi\hbar^2(3a_4-10a_2+7a_0)/7M_\text{a}$ are derived from the $s$-wave scattering lengths $a_s$ for the total spin $s$ of the interacting atom pairs, $a_0=87.4(10)a_\mathrm{B}$, $a_2=92.4(10)~a_\mathrm{B}$, and $a_4=100.5(10)~a_\mathrm{B}$ in units of the Bohr magneton $a_\mathrm{B}$~\cite{klausen_pra_2001}.

For a weak quadratic Zeeman shift, the spin-2 BEC exhibits different magnetic phases that are characterized by the values of $|\langle\Fhat\rangle|$ and $|A_{20}|$. For a ferromagnetic phase, $|A_{20}|=0$ and $|\langle\Fhat\rangle|=2$ (for FM2) or 1 (for FM1). The representative spinors are obtained by rotations of $(1,0,0,0,0)^T$ and $(0,1,0,0,0)^T$, respectively. For the C phase, $|\langle\Fhat\rangle|=|A_{20}|=0$ and the spinor representations are obtained by rotations of $\tfrac{1}{\sqrt{3}}(1,0,0,\sqrt{2},0)^T$. For spin-2 BECs, the polar magnetic phase, which also has a vanishing spin magnitude $|\langle\Fhat\rangle|=0$ but
$|A_{20}|=\frac{1}{\sqrt{5}}$, is found to separate into two distinct phases: UN and BN, represented by the spinors $(0,0,1,0,0)^T$ and $\tfrac{1}{\sqrt{2}}(1,0,0,0,1)^T$, respectively. By defining the
amplitude of spin-singlet trio formation~\cite{kawaguchi_physrep_2012}
$A_{30}=\frac{3\sqrt{6}}{2}(\zeta_{+1}^2\zeta_{-2}+ \zeta_{-1}^2\zeta_{+2}) + \zeta_0 (
-6\zeta_{+2}\zeta_{-2} -3\zeta_{+1}\zeta_{-1}+\zeta_0^2)$ we can distinguish between the UN and BN phases that take the values $|A_{30}|=1$ and 0, respectively. For the C phase $|A_{30}|=\sqrt{2}$.

We solve the five coupled Gross-Pitaevskii equations obtained from Eq.~\eqref{ham} numerically using a split-step method~\cite{javanainen_jpa_2006} and experimental parameter values. Angular momentum is conserved in $s$-wave scattering, implying that on time scales where this dominates, longitudinal magnetization, $M_z = (1/N)\int \mathrm{d}^3r\, n(\rr) F_z(\rr)$, where $N$ is the total number of atoms, is conserved. We employ an algorithm that explicitly conserves $M_z$.

\subsection{Graphical illustration of magnetic phase symmetries} We represent the order-parameter symmetry of the magnetic phases in terms of the complementary spherical-harmonics and Majorana representations.
The stars of the Majorana constellation~\cite{majorana_nuovocimento_1932} are calculated
by numerically finding the $2F+1$ roots $z=z_j$ of the complex polynomial,
\begin{align}
\sum_{\alpha=0}^{2F}\sqrt{\begin{pmatrix}2F \\ \alpha \end{pmatrix}}\,\zeta_{F-\alpha}^* z^\alpha = 0,
\end{align}
where each root then represents a stereographic projection $z_j = \tan\left( \frac{\theta}{2}\right)e^{i\phi}$ of the spherical coordinates $(\theta,\phi)$ that define the Majorana points.
Rotations of the state are equivalent to rotations of the constellation on the Majorana sphere that preserve their relative orientation. In the spherical harmonics representation we show $|Z(\theta,\phi)|^2$, where $Z(\theta,\phi)=\sum_{m=-F}^{+F}Y_{F,m}(\theta,\phi)\zeta_m$ expands the state in terms of the spherical harmonics $Y_{F,m}(\theta,\phi)$ ($F=1,2$). The symmetry is completed by the color scale that is obtained from $\mathrm{Arg}(Z)$.

\subsection{Magnetic phases and singular vortex lines} The spin-2 FM2/1 and UN phases are closely related to the spin-1 ferromagnetic (FM) and polar (P) phases, where the order parameter of the FM phase corresponds to the group of spatial rotations, SO(3). The singular line defects in such a system, as in the FM1 phase of spin-2, can only belong to two topologically distinct equivalence classes, representing the winding numbers zero and one~\cite{weiss_ncomm_2019}. The FM2 phase in the spin-2 system doubles the number of these classes to four. The UN phase is determined by the nematic axis $\hat {\bf d}$ and the global phase of the macroscopic condensate wave function $\tau$.
The order parameter symmetry is $S^1\times S^2/\mathbb{Z}_2$, where the two-element group factorization is due to the identity of the states $\zeta(\hat {\bf d},\tau)=\zeta(-\hat {\bf d},\tau) $. This nematic symmetry allows for the existence of spin half-quantum vortices (HQVs).

The BN and C phases exhibit polytope symmetries, as shown in Fig.~\ref{fig:phases}f--g, that result in a much richer structure of singular line defects.
The topologically distinct families of line defects in such magnetic phases derive from the conjugacy classes of the fundamental homotopy group $\pi_1$ of the corresponding order-parameter space symmetries. Specifically, these are determined directly from the group of transformations that leave the order parameter unchanged~\cite{mermin_rmp_1979}---here by applying a $\SO(3)$ spin rotation and a $S^1$ gauge transformation by the global condensate phase $\tau$. For the case of the BN phase, the fourfold symmetry of the order parameter is illustrated in Fig.~\ref{fig:phases}f. Transformations that exactly interchange the lobes in the spherical-harmonics representation and also, where necessary, take $\tau \to \tau+\pi$, then leave the order parameter unchanged. These transformations thus combine the dihedral-4 subgroup of $\SO(3)$ with a $\pi$ shift of $\tau$ to form the eight-element group
$\Dt_4$, which factorizes $S^1\times\SO(3)$. After lifting $\SO(3)$ to $\SU(2)$ to form a simply-connected covering group, the conjugacy classes of $\pi_1$ are obtained~\cite{borgh_prl_2016}: $\{(n,\mathbf{1})\}$,
$\{(n,-\mathbf{1})\}$,
$\{(n,\pm i\sx), (n,\pm i\sy)\}, \{(n, \pm i\sz)\}$,
$\{(n+1/2,\sigma), (n+1/2,-i\sz\sigma)\}$,
$\{(n+1/2,-\sigma), (n+1/2,i\sz\sigma)\}$, and
$\{(n+1/2,\pm i\sx\sigma),(n+1/2,\pm i\sy\sigma)\}$,
where the Pauli matrices $\sigma_{x,y,z}$ and
$\sigma\equiv\left(\mathbf{1}+i\sigma_z\right)/\sqrt{2}$ represent the
$\SU(2)$ part of the fundamental homotopy group elements, and $n$ in the $S^1$ part
is an integer.
For $n=0$, these represent the following topologically distinguishable vortex states: $(i)$~the vortex-free state,
$(ii)$~integer spin vortex, $(iii)$ \& $(iv)$~spin HQVs,
$(v)$~HQV with $\pi/2$ spin rotation,
$(vi)$~HQV with $3\pi/2$ spin rotation, and
$(vii)$~HQV with $\pi$ spin rotation. For $n=1$, the singly-quantized vortices in Fig.~\ref{fig:leveldiagram}a,b arise from $(i)$.

The vortex classes in the C phase are determined through an analogous analysis~\cite{semenoff_prl_2007} by noting that the order parameter is left invariant by transformations in the 12-element group $\tilde{T}$ that combines the tetrahedral subgroup of $\SO(3)$ with elements of $S^1$. This reveals the vortex classes
$\{(n,\mathbf{1})\}$,
$\{(n,-\mathbf{1})\}$,
$\{(n,\pm i\sigma_{\alpha)}\}$,
$\{(n+1/3,\bar{\sigma}), (n+1/3,-i\sigma_{\alpha}\bar{\sigma})\}$,
$\{(n+1/3,-\bar{\sigma}), (n+1/3,i\sigma_{\alpha}\bar{\sigma})\}$,
$\{(n+2/3,\bar{\sigma}^2),(n+2/3,-i\sigma_{\alpha}\bar{\sigma}^2)\}$, and $\{(n+2/3,-\bar{\sigma}^2),(n+2/3,i\sigma_{\alpha}\bar{\sigma}^2)\}$, where $\alpha=x,y,z$ and $\bar{\sigma}\equiv\tfrac{1}{2}\left(\mathbf{1}+i\sigma_x+i\sigma_y+i\sigma_z\right)$. For $n=0$, the latter four classes represent vortices with fractional $\tfrac{1}{3}$ and $\tfrac{2}{3}$ charges, while singly quantized vortices arise from the first three with $n=1$.


\section{Experimental procedures}
The experiment begins with condensates of approximately $2\times 10^5$ atoms in the $\ket{F=1,m_F=1}$ spinor component of $^{87}$Rb. The BEC is confined in an optical trap with frequencies $(\omega_r ,\omega_z) \approx 2\pi(130,170)~\text{s}^{-1}$. Three pairs of Helmholtz coils generate the magnetic bias field, and a single anti-Helmholtz pair generates the magnetic quadrupole field. 

The condensate is initially in a magnetic field of $17$~mG along the $+z$-axis and a radial magnetic gradient of strength $4.3$~G/cm. The field vanishes at a point approximately $20~\micron$ above the condensate, and the spins initially point along the $+z$-axis. The bias field is then reduced at a rate of 5~G/s until it reaches $-57$~mG, drawing the zero point of the magnetic field through the condensate and imprinting the nonsingular vortex as described in the main text. Subsequently, the gradient is quenched and the field is quickly reoriented to point once again along the $+z$ axis at $1.00$~G for at least 100~ms to allow the vortex to decay into two singular SO(3) vortices in the FM phase with filled P cores due to a very sharp bending of the spin texture~\cite{lovegrove_pra_2016,weiss_ncomm_2019}. Field drifts are slow enough to permit adjustment to the bias fields to ensure that the magnetic field zero passes through the approximate center of the condensate.

Microwave and rf pulses are applied through a waveguide and a single coil oriented along the $y$-axis, respectively, and are initiated synchronously with the 60~Hz power line frequency to minimize the effects of ambient time-dependent magnetic fields. A typical sequence contains more than five resonant pulses and lasts 200-400$~\mu$s, which is short compared with the typical evolution time of the condensate, $\sim (c_0 n)^{-1}$. Control of the magnetic fields at the milligauss level is essential, and care must be taken in the design of the pulse sequence to ensure that unwanted transitions are not driven inadvertently because of frequency-broadening and degeneracies in the hyperfine level structure.
Examples of these pulse sequences are given in Fig.~\ref{fig:leveldiagram} for BN vortex with UN core and in the Supplemental Material~\cite{supplementalmaterial} for BN vortex with FM2 core, for C vortex with FM2 core, and for C vortex with UN core.

To image the condensates, we turn off the optical trap, whereupon the condensate falls freely and expands ballistically. A briefly applied magnetic field gradient shortly after release separates the spinor components for simultaneous absorption imaging along the horizontal and vertical axes, which provides a pair of five atomic column density profiles for $m \in \{-2,-1,0,1,2\}$. The incident probe beam is closest to resonance for $m=+2$ atoms to the optically excited states, and therefore the imaging efficiency of the spin-2 magnetic states decreases as the magnetic number $m$ decreases. We have calibrated this effect and normalized the atomic density of the condensate spinor components accordingly for all images and analysis in this work.

\begin{figure*}
\centering
\includegraphics[width=\linewidth]{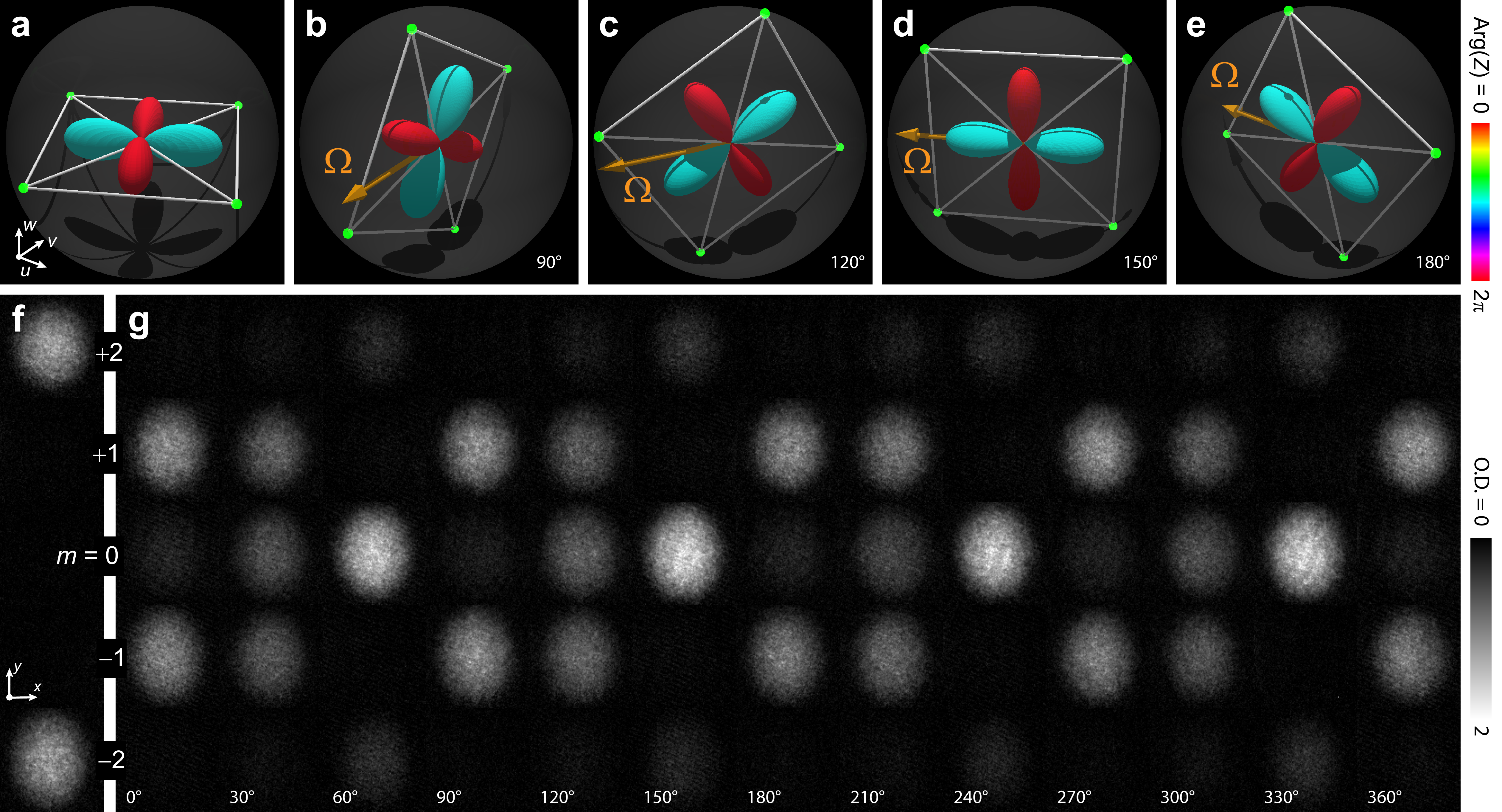}
\caption{\textbf{Demonstration of polytope discrete symmetry in the continuous wave function of the spin-2 superfluid in the BN phase.} The different orientations of the order parameter are obtained by rotations of the spinor with radio-frequency (rf) pulses at different phase angles, as shown schematically by the orange torque vector $\boldsymbol{\Omega}$. \textbf{a,} Majorana and spherical harmonics representations of the initial BN magnetic phase. \textbf{b--e,} The rotated BN phase after $\pi/2$ rf pulses with the phases $90^\circ$, $120^\circ$, $150^\circ$, and $180^\circ$, as shown. \textbf{f,} Experimental images of the spinor components for the initial magnetic phase. \textbf{g,} Experimental images of the rotated spinor components after $\pi/2$ rotations at the indicated rf phases. Each experimental subpanel shows a spinor component column density taken from the side in terms of optical depth (O.D.) with a field of view of $219~\micron \times 219~\micron$. \label{fig:bnphase}}
\end{figure*}

\begin{figure*}
\centering
\includegraphics[width=\linewidth]{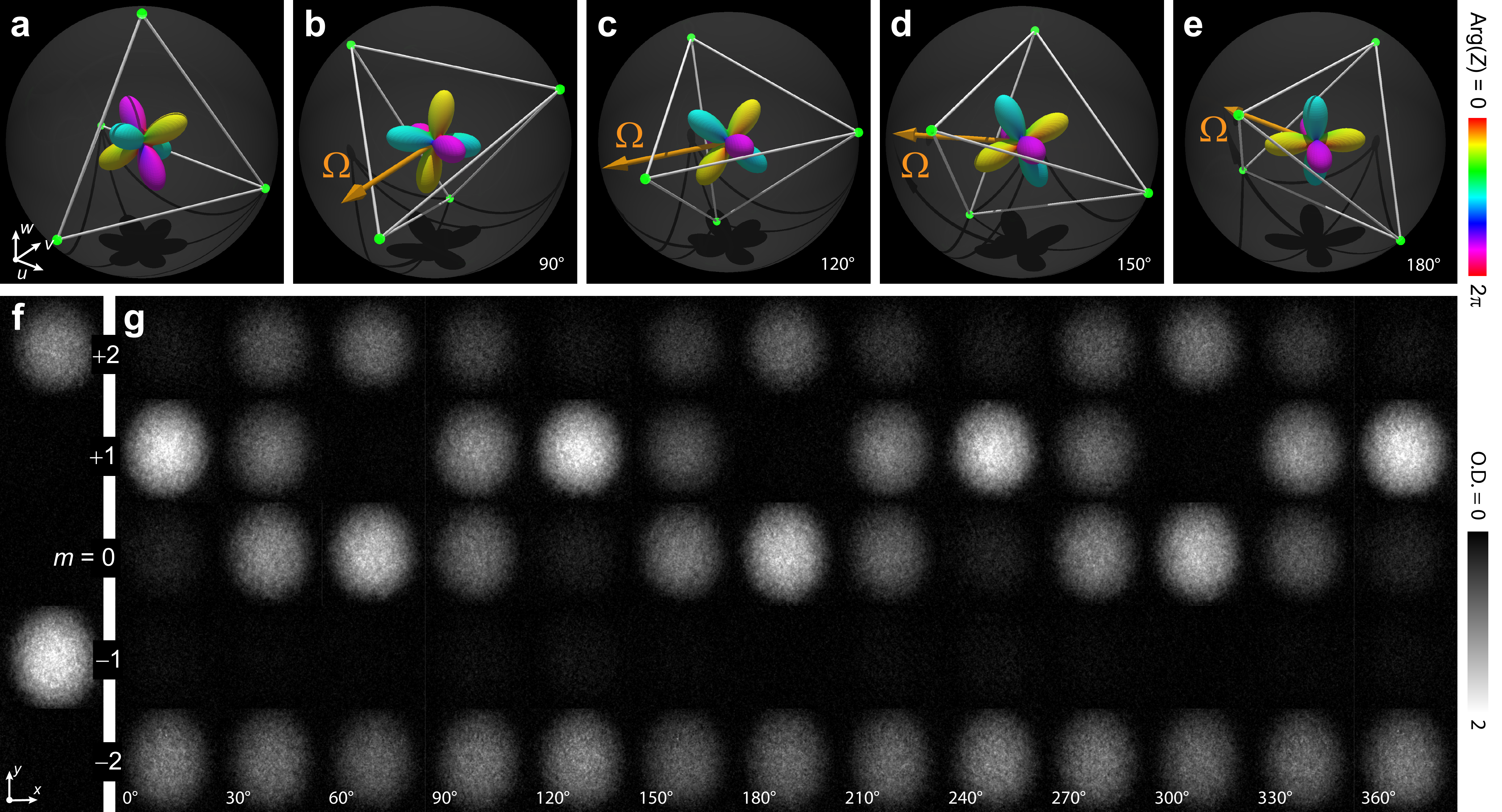}
\caption{\textbf{Demonstration of polytope discrete symmetry in the continuous wave function of the spin-2 superfluid in the C phase.} The different orientations of the order parameter are obtained by rotations of the spinor with rf pulses at different phase angles, as shown schematically by the orange torque vector $\boldsymbol{\Omega}$. \textbf{a,} Majorana and spherical harmonics representations of the initial C magnetic phase. \textbf{b--e,} The rotated C phase after tetrahedral angle rf pulses with the phases $90^\circ$, $120^\circ$, $150^\circ$, and $180^\circ$, as shown. \textbf{f,} Experimental images of the spinor components for the initial magnetic phase. \textbf{g,} Experimental images of the rotated spinor components, taken from the side after $\pi/2$ rotations at the indicated rf phases. Each experimental subpanel shows a spinor component column density taken from the side in terms of optical depth (O.D.) with a field of view of $219~\micron \times 219~\micron$.\label{fig:cphase}}
\end{figure*}

In order to explicitly demonstrate the discrete polytope point group symmetries, we control the orientation of the order parameters by inducing a spin rotation. The rotation axis and angle of rotation are determined by the phase and area of the rf pulse.
We use a pair of phase-locked direct digital synthesis (DDS) rf sources to control the frequency and phase of the applied rf and microwave signals. One source ($\sim0.7$~MHz) directly generates the rf spin-tip pulses; the other ($\sim30$~MHz) is mixed with a carrier ($6804$~MHz) to generate the microwaves. Reproducible phase adjustments are achieved by synchronizing the phase offset of both DDS sine waves at a fixed time before the pulse sequence begins. Selection of the phase for the rf spin-tip pulse achieves a rotation of the Majorana constellation about the corresponding axis in the co-rotating $uv$-plane defined in Fig.~\ref{fig:un_phases}a--d. Examples of full $2\pi$ rotations of the rf phase, demonstrating the discrete polytope symmetries of the BN and C phases, are shown in Figs.~\ref{fig:bnphase} and~\ref{fig:cphase}, respectively.


%

\clearpage
\pagebreak

\setcounter{page}{1}
\setcounter{figure}{0}
\renewcommand{\thepage}{SM--\arabic{page}}

\setcounter{figure}{0}
\renewcommand*{\thefigure}{SM\arabic{figure}}

\onecolumngrid

\begin{center}
\Huge Supplemental Material for
\end{center}

\begin{center}
\Large \textbf{Topological Superfluid Defects with Discrete Point Group Symmetries}
\end{center}

\begin{center}
\normalsize Y. Xiao, M.O.\ Borgh, A.A.\ Blinova, T. Ollikainen, J. Ruostekoski, and D.S.\ Hall
\end{center}

\addvspace{0.5in}

\begin{figure*}[hb!]
\centering
\includegraphics[width=0.5\linewidth]{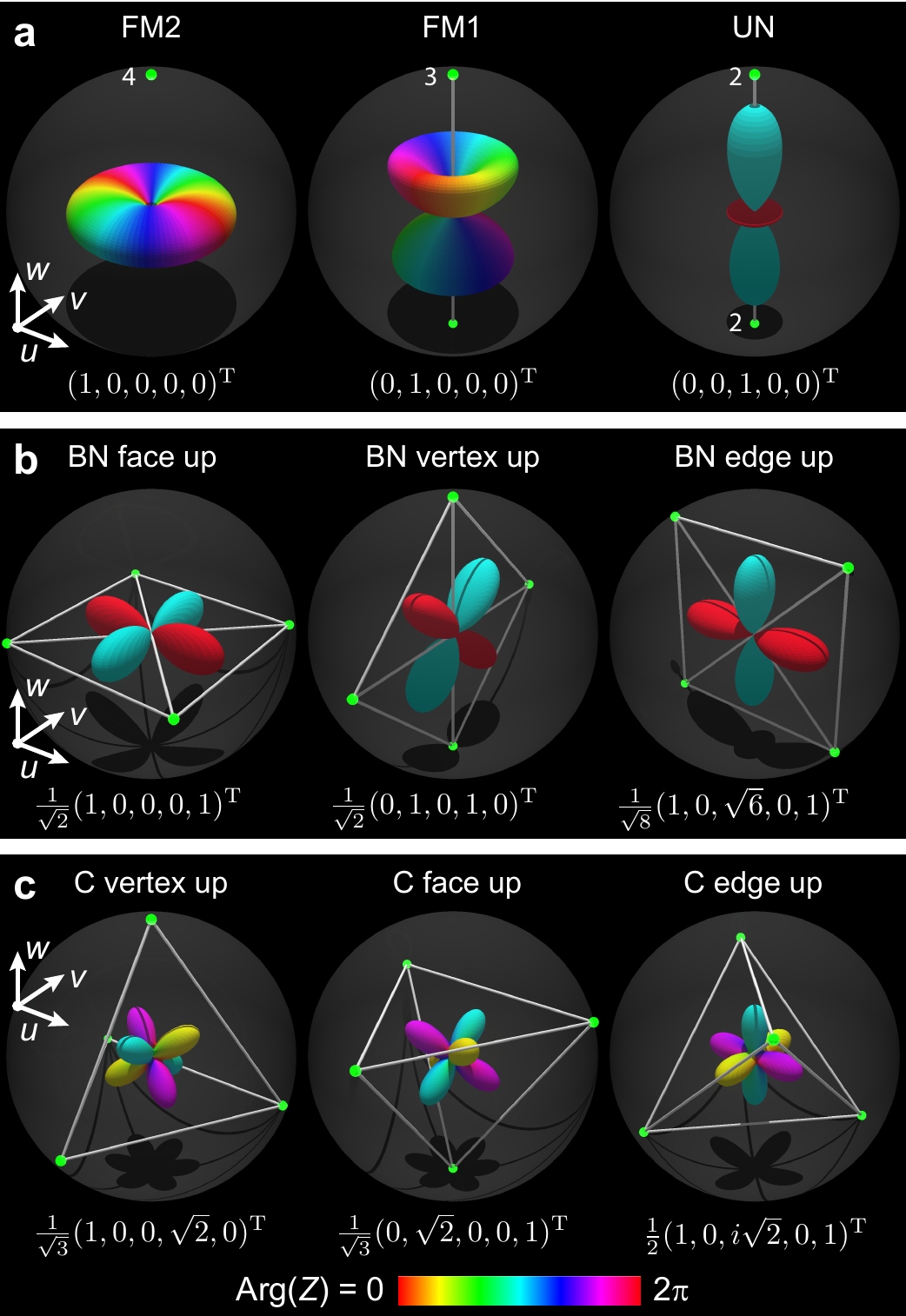}
\caption{\textbf{Magnetic phase symmetries of spin-2 Bose--Einstein condensates.}  \textbf{a,} Prototype spinors, Majorana and spherical harmonics representations of the magnetic phases without discrete polytope symmetry: ferromagnetic-2 (FM2), ferromagnetic-1 (FM1), and uniaxial nematic (UN). \textbf{b,} The representations of the biaxial nematic (BN) phase with the Majorana symmetry of a square, for three different orientations. \textbf{c,} The representations of the cyclic (C) phase with the Majorana symmetry of a tetrahedron, for three different orientations. The Majorana points are shown as green dots with adjacent numbers indicating multiplicities $>1$.
\label{sfig:bncorientations}}
\end{figure*}

\begin{figure*}
\centering
\includegraphics[width=0.7\linewidth]{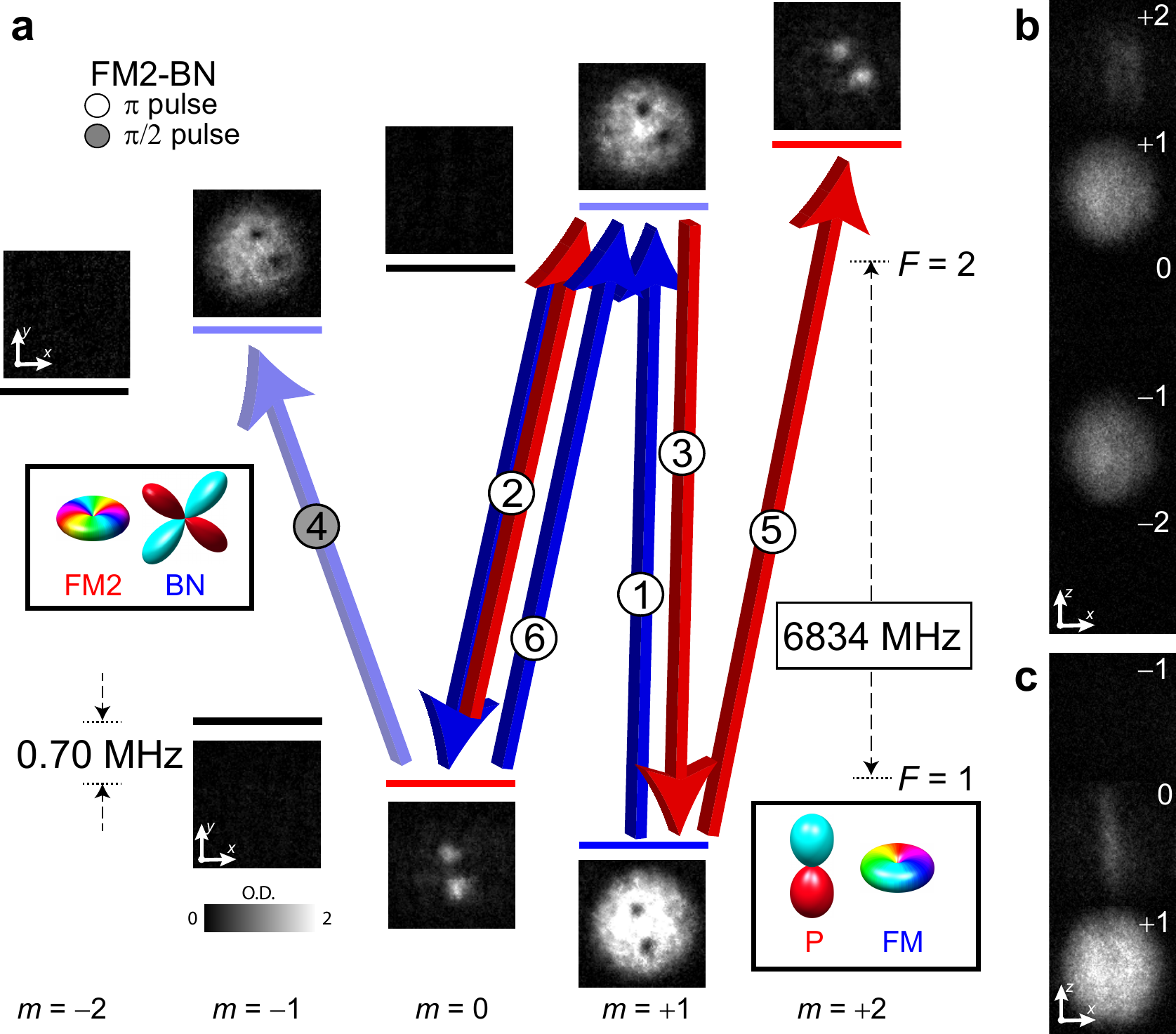}
\caption{\textbf{Creation of singular vortices in the BN phase with FM2 cores.} \textbf{a,} The circled numbers give the order of the microwave transitions within the hyperfine levels (solid lines), beginning with a vortex in the spin-1 ferromagnetic (FM) phase with polar (P) core in the $|F,m\rangle$ states shown. The red arrows indicate the path of the cores through the internal states of the system, and the blue arrows indicate the path of the vortices. In the fourth pulse (pale blue arrow), the $\pi/2$ rotation angle transfers half of the population from the $|F=1,m=0\rangle$ spinor component to the $|2,+1\rangle$ component. The experimental images show column densities taken from (a) the top, and the side for \textbf{b,} $F=1$ and \textbf{c,} $F=2$, expressed in units of optical depth (O.D.) with a field of view of $219~\micron \times 219~\micron$. The images from the upper and lower hyperfine levels are from different condensates.\label{sfig:F2-BNP}}
\end{figure*}

\begin{figure*}
\centering
\includegraphics[width=0.7\linewidth]{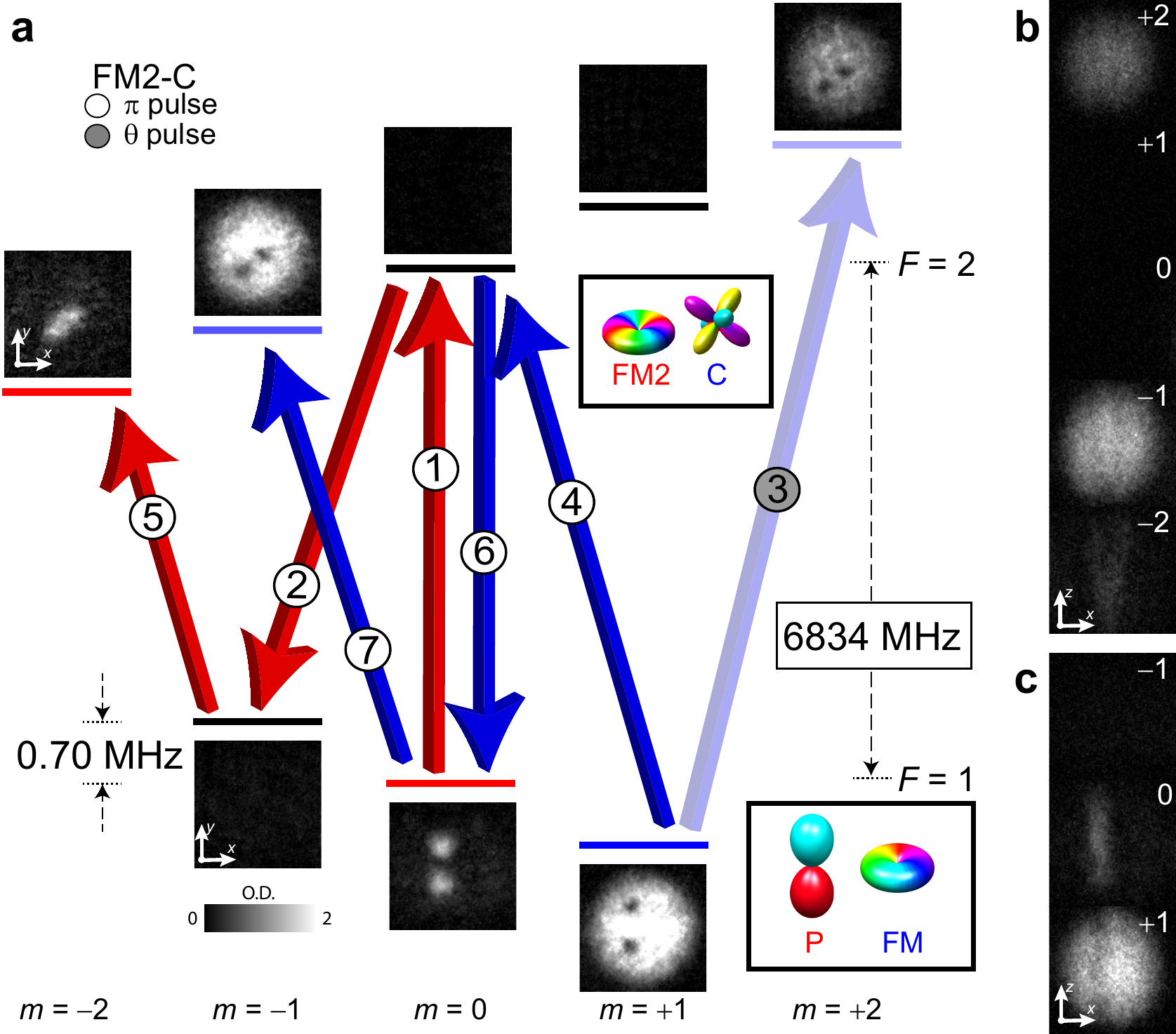}
\caption{\textbf{Creation of singular vortices in the C phase with FM2 cores.} \textbf{a,} The thick lines schematically show the hyperfine levels ($F=1$ and $F=2$, with Zeeman levels $m$ reading left to right) in a magnetic field of 1~G, accompanied by experimental images of the condensate viewed along the $z$-axis after Stern--Gerlach separation. The connecting arrows illustrate the pulse sequence, with order given by the circled number, colored blue for transitions involving the components with phase singularities and red involving the superfluid components filling those singularities. The sequence begins with a vortex in the spin-1 FM phase with P core. In the third pulse (pale blue arrow), the rotation angle $\theta=2\arcsin(1/\sqrt{3})$ transfers 1/3 of the population from the $|F=1,m=+1\rangle$ spinor component to the $|2,+2\rangle$ component. The experimental images show column densities taken from (a) the top, and the side for \textbf{b,} $F=1$ and \textbf{c,} $F=2$, expressed in units of optical depth (O.D.) with a field of view of $219~\micron \times 219~\micron$. The images from the upper and lower hyperfine levels are from different condensates.\label{sfig:F2-C}}
\end{figure*}

\begin{figure*}
\centering
\includegraphics[width=0.64\linewidth]{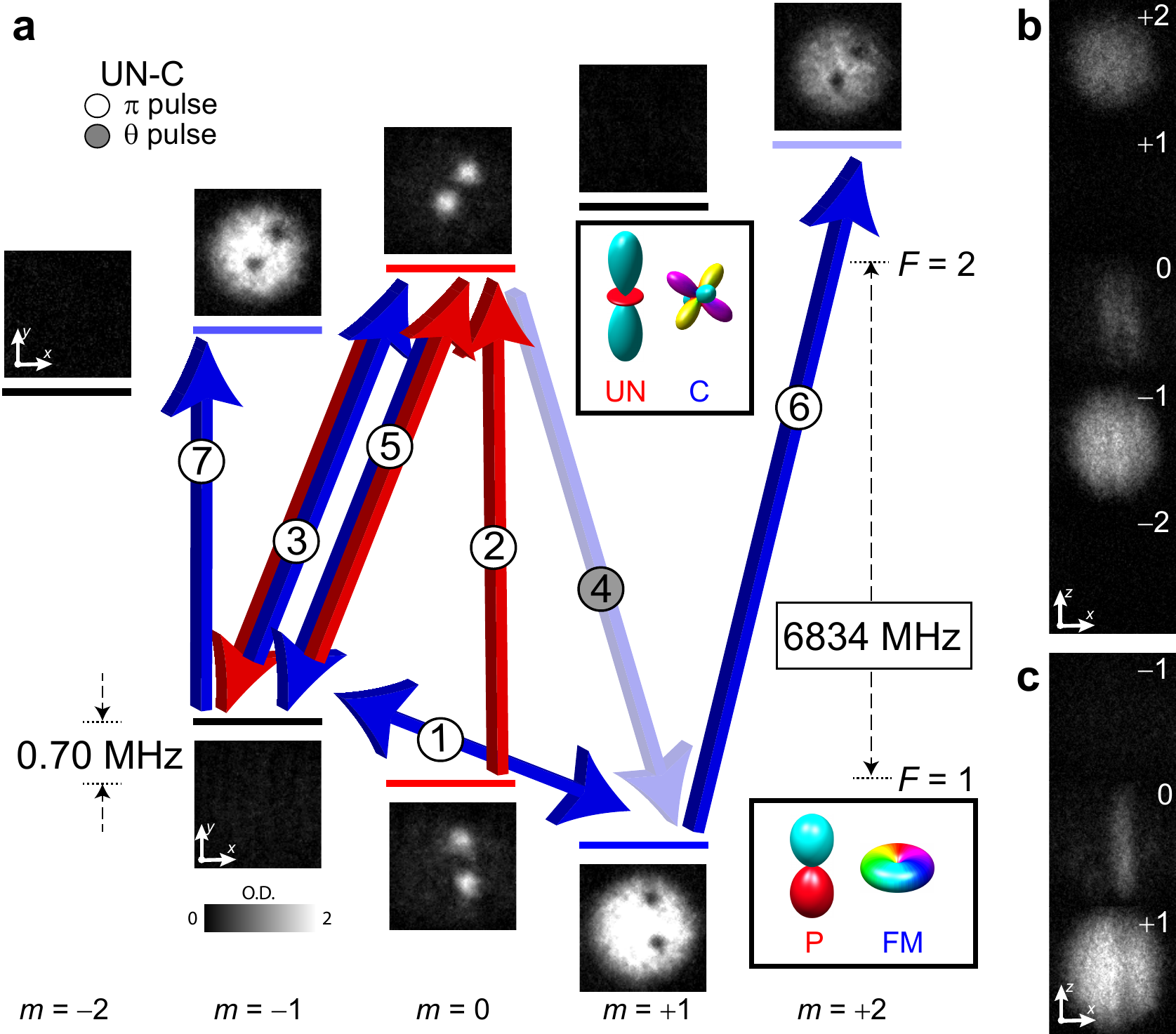}
\caption{\textbf{Creation of singular vortices in the C phase with UN cores.} \textbf{a,} The thick lines schematically show the hyperfine levels ($F=1$ and $F=2$, with Zeeman levels $m$ reading left to right) in a magnetic field of 1~G, accompanied by experimental images of the condensate viewed along the $z$-axis after Stern--Gerlach separation. The connecting arrows illustrate the pulse sequence, with order given by the circled number, colored blue for transitions involving the components with phase singularities and red involving the superfluid components filling those singularities. The sequence begins with a vortex in the spin-1 FM phase with P core. The first rf $\pi$-pulse converts the $|F=1,m=+1\rangle$ spinor component into the $|1,-1\rangle$ component, leaving the $m=0$ component unchanged; a similar sequence, omitting this rotation, can create the face-up C vortex with UN core. In the fourth pulse (pale blue arrow), the rotation angle $\theta=2\arcsin(1/\sqrt{3})$ transfers 1/3 of the population from the $|2,0\rangle$ spinor component to the $|1,+1\rangle$ component. The experimental images show column densities taken from (a) the top, and the side for \textbf{b,} $F=1$ and \textbf{c,} $F=2$, expressed in units of optical depth (O.D.) with a field of view of $219~\micron \times 219~\micron$. The images from the upper and lower hyperfine levels are from different condensates.\label{sfig:UN-C}}
\end{figure*}

\end{document}